\documentclass[12pt]{article}
\usepackage{amssymb}

\input{tcilatex}

\begin{document}

\bigskip

\bigskip\ 

\begin{center}
\textbf{THE DE SITTER RELATIVISTIC TOP THEORY}

\bigskip\ 

\smallskip\ 

J. Armenta$^{a}$ and J. A. Nieto$^{b}$\footnote[1]{%
nieto@uas.uasnet.mx}

\smallskip\ 

\smallskip\ 

$^{a}$\textit{Departamento de Investigaci\'{o}n en F\'{\i}sica de la
Universidad de Sonora,}

\textit{83000 Hermosillo Sonora, M\'{e}xico}

$^{a}$\textit{Instituto Tecnol\'{o}gico y de Estudios Superiores de}

\textit{Monterrey, Campus Obreg\'{o}n, Apdo. Postal 622, 85000 Cd. Obreg\'{o}%
n Sonora, M\'{e}xico}

$^{b}$\textit{Facultad de Ciencias F\'{\i}sico-Matem\'{a}ticas de la
Universidad Aut\'{o}noma}

\textit{de Sinaloa, 80010 Culiac\'{a}n Sinaloa, M\'{e}xico}

\bigskip\ 

\bigskip\ 

\textbf{Abstract}
\end{center}

We discuss the relativistic top theory from the point of view of the de
Sitter (or anti de Sitter) group. Our treatment rests on Hanson-Regge's
spherical relativistic top lagrangian formulation. We propose an alternative
method for studying spinning objects via Kaluza-Klein theory. In particular,
we derive the relativistic top equations of motion starting with the
geodesic equation for a point particle in $4+N$ dimensions. We compare our
approach with the Fukuyama's formulation of spinning objects, which is also
based on Kaluza-Klein theory. We also report a generalization of our
approach to a $4+N+D$ dimensional theory.

\bigskip\ 

\bigskip\ 

\bigskip\ 

\bigskip\ 

\bigskip\ 


Pacs numbers: 04.20.-q, 04.60.+n, 11.15.-q, 11.10.Kk

August, 2004

\newpage

\noindent \textbf{1.- INTRODUCTION}

\smallskip\ 

If one compares the Regge's references 1 and 2 published in 1959 and 1960
respectively with the Hanson-Regge's reference of 1974 about the
relativistic spherical top theory$^{3}$ (see also Ref. 4), one gets the
feeling that Regge thought on the trajectory constraint linking mass and the
spin of a relativistic spinning object as a deep physical concept of nature.
Through the years it has become clear that Regge was right. In fact, such a
constraint, now called Regge trajectory, plays a fundamental role not only
in the dual string models$^{5}$ and the relativistic rotator theory,$^{6-7}$
but also in string theory$^{8}$ and in black holes approach.$^{9}$ It seems
that even Regge is in the sky$^{10}$ in connection with the mass and the
internal angular momentum of celestial objects.

One of the simplest Regge trajectory for a spherical relativistic top is
provided by the expression;$^{3-4}$

\begin{equation}
H\equiv P^{\mu }P_{\mu }+\frac{1}{2r^{2}}\Sigma ^{\mu \nu }\Sigma _{\mu \nu
}+m_{0}^{2}\thickapprox 0,  \tag{1}
\end{equation}%
where $P^{\mu }$ and $\Sigma ^{\mu \nu }=-\Sigma ^{\nu \mu }$ are the linear
momentum and the internal angular momentum respectively associated with some
spinning object. Here, $m_{0}$ and $r$ are constants determining the
properties of the system and the symbol "$\thickapprox 0$" means weakly
equal zero in the sense of the terminology of Dirac's constraints
Hamiltonian formalism.$^{11}$ (Here the indices $\mu ,\nu $ run from 0 to 3.)

One of the interesting aspects of (1) is that resembles one of the Casimir
operator of the de Sitter group

\begin{equation}
C_{1}=\frac{1}{2}S^{AB}S_{AB},  \tag{2}
\end{equation}%
where the indices $A,B$ run from $0$ to $4$. In fact, if classically it is
possible to make the identifications

\begin{equation}
S^{4\mu }\rightarrow rP^{\mu },S^{\mu \nu }\rightarrow \Sigma ^{\mu \nu }%
\text{ and }C_{1}\rightarrow -r^{2}m_{0}^{2},  \tag{3}
\end{equation}%
then the spherical relativistic top may lead naturally to a de Sitter
relativistic top and several properties of de Sitter group can be applied to
such a system. The problem, however, it is not so simple because the momenta 
$P^{\mu }$ and $\Sigma ^{\mu \nu }$ are restricted to satisfy the so called
Tulczyjew constraint$^{12}$

\begin{equation}
\Sigma ^{\mu \nu }P_{\nu }\thickapprox 0  \tag{4}
\end{equation}%
and it seems that there is not a counterpart in the de Sitter group
formalism of this constraint. One of the main goals of this work is to use a
lagrangian analysis of the relativistic top in order to shed some light on
this and other related problems.

As soon as we make the identification (3) the parameter $r,$ measuring the
"size"\ of the top, may acquire a particular interesting meaning, namely, it
can be identified with the Planck length $l_{P}=(\hslash G/c^{3})^{1/2}$ or
with the radius of the universe $R$. In the first case, the relativistic top
may have contact not only with elementary particles through the
superstringtop theory,$^{13-14}$ but also with gravity itself.$^{15}$ In
fact, it has been shown$^{15}$ that extending the Poincar\'{e} group to the
de Sitter group through a Wigner contraction with $l_{P}$ as a parameter one
can make sense of a gravitational theory as a gauge theory. Similar
conclusion is provided by the MacDowell-Mansouri formalism.$^{16}$

In the second case, one may find a connection between the relativistic top
with accelerated universe via the de Sitter spacetime. In some sense, one
may say that Regge is not only in the sky but in the cosmos as well.

The central idea of this work is to develop different aspects of the de
Sitter top theory using the spherical relativistic top theory as a guide.
For this purpose in sections 2 and 3 we show explicitly how the first order
and second order formalisms of a particular spherical top system are
related. In section 4, we show how the Kaluza-Klein formalism may lead to de
Sitter top theory. In section 5, we make some final remarks. Finally, in the
appendix B, we report a generalization of our formalism to $4+N+D$
dimensions.

\bigskip\ 

\noindent \textbf{2.- FROM THE FIRST ORDER TO THE SECOND ORDER LAGRANGIAN}

\smallskip\ 

Let us describe the motion of the top by four coordinates $x^{\mu }(\tau )$
and a tetrad $e_{(\alpha )}^{\mu }(\tau )$ where $\tau $ is an arbitrary
parameter along the world line of the top. Here $x^{\mu }(\tau )$ is used to
describe the position of the system, while $e_{(\alpha )}^{\mu }(\tau )$ is
attached to the top in order to describe its rotations. We shall assume that
the tetrad $e_{(\alpha )}^{\mu }(\tau )$ satisfies the orthonormal relations:

\begin{equation}
\begin{array}{c}
\eta _{\mu \nu }e_{(\alpha )}^{\mu }e_{(\beta )}^{\nu }=\eta _{(\alpha \beta
)}, \\ 
\\ 
\eta ^{(\alpha \beta )}e_{(a)}^{\mu }e_{(\beta )}^{\nu }=\eta ^{\mu \nu },%
\end{array}
\tag{5}
\end{equation}
where $\eta _{\mu \nu }=diag(-1,1,1,1)$ is the Minkowski metric. We shall
associate the canonical linear momentum $P_{\mu }$ to the linear velocity $%
u^{\mu }=\frac{dx^{\mu }}{d\tau }$ and the spin tensor $\Sigma ^{\mu \nu
}=-\Sigma ^{\nu \mu }$ to the angular velocity $\sigma ^{\mu \nu
}=e_{(\alpha )}^{\mu }\frac{de_{(\alpha )}^{\nu }}{d\tau }.$

Consider the first order lagrangian corresponding to a special type of a
relativistic top:$^{3-4}$

\begin{equation}
L=u^{\mu }P_{\mu }+\frac{1}{2}\sigma ^{\mu \nu }\Sigma _{\mu \nu }-\frac{%
\lambda }{2}\left( P^{\mu }P_{\mu }+\frac{1}{2r^{2}}\Sigma ^{\mu \nu }\Sigma
_{\mu \nu }+m_{0}^{2}\right) +\xi _{\mu }\Sigma ^{\mu \nu }P_{\nu },  \tag{6}
\end{equation}%
where $\lambda $ and $\xi _{\mu }$ are lagrange multipliers. Varying $L$
with respect to $P_{\mu }$ gives

\begin{equation}
u^{\mu }=\lambda P^{\mu }+V^{\mu },  \tag{7}
\end{equation}%
where%
\begin{equation}
V^{\mu }=\Sigma ^{\mu \nu }\xi _{\nu },  \tag{8}
\end{equation}%
while varying $L$ with respect to $\Sigma _{\mu \nu }$ yields

\begin{equation}
\sigma ^{\mu \nu }=\frac{\lambda }{r^{2}}\Sigma ^{\mu \nu }+P^{\mu }\xi
^{\nu }-P^{\nu }\xi ^{\mu }.  \tag{9}
\end{equation}

Similarly varying $L$ with respect to $\xi _{\mu }$ leads to the constraint

\begin{equation}
\Sigma ^{\mu \nu }P_{\nu }=0,  \tag{10}
\end{equation}%
while varying $L$ with respect to $\lambda $ one obtains

\begin{equation}
P^{\mu }P_{\mu }+\frac{1}{2r^{2}}\Sigma ^{\mu \nu }\Sigma _{\mu \nu
}+m_{0}^{2}=0.  \tag{11}
\end{equation}%
Since $\Sigma ^{\mu \nu }$ is antisymmetric, we observe from (6) that if $%
\xi _{\mu }$ is parallel to $P_{\mu }$ the last term in (6) vanishes
identically. Therefore, we may set

\begin{equation}
\xi ^{\mu }P_{\mu }=0.  \tag{12}
\end{equation}%
This condition must be added to the lagrangian (6) in the form $\eta (\xi
^{\mu }P_{\mu })$, where $\eta $ is another lagrange multiplier. Varying the
resultant extended lagrangian with respect to $\eta $ leads to (12). While
arbitrary variations with respect to $\xi ^{\mu }$ leads to the equation $%
\Sigma ^{\mu \nu }P_{\nu }+\eta P^{\mu }=0$ which, after multiplying it by $%
P_{\mu },$ gives $\eta =0$ and therefore one recovers the lagrangian (6).

Our goal is to derive the second order lagrangian associated with the
lagrangian (6). Our proof consists of some elementary algebra and for that
reason in this section we shall only mention the main results. Nevertheless,
since in such an algebra there are some key steps, in the appendix A we
present the computation in more detail.

The main idea is to compute from (7)-(12) the combination $u\sigma \sigma
u-r^{2}\det \sigma $, where

\begin{equation}
u\sigma \sigma u\equiv u_{\mu }\sigma ^{\mu \nu }\sigma _{\nu }{}^{\alpha
}u_{\alpha }  \tag{13}
\end{equation}%
and

\begin{equation}
\det \sigma =\frac{1}{4!}\varepsilon ^{\mu \nu \alpha \beta }\varepsilon
^{\tau \lambda \sigma \rho }\sigma _{\mu \tau }\sigma _{\nu \lambda }\sigma
_{\alpha \sigma }\sigma _{\beta \rho }.  \tag{14}
\end{equation}%
One finds the following result (for details see appendix A);

\begin{equation}
u\sigma \sigma u-r^{2}\det \sigma =\frac{m_{0}^{2}}{r^{2}}\left[ \left(
u^{2}+\frac{r^{2}\sigma ^{2}}{2}\right) \lambda ^{2}+\frac{1}{r^{2}}\lambda
^{4}m_{0}^{2}\right] ,  \tag{15}
\end{equation}%
where we used the notation $a^{2}=a^{\mu }a_{\mu },$ for any dynamical
variable $a^{\mu }$. Using once again the constraint (11) we find that (15)
leads to the expression

\begin{equation}
\lambda ^{4}+\frac{\lambda ^{2}}{m_{0}^{2}}\left( u^{2}+\frac{1}{2}%
r^{2}\sigma ^{2}\right) -\frac{r^{2}}{m_{0}^{4}}\left( u\sigma \sigma
u-r^{2}\det \sigma \right) =0.  \tag{16}
\end{equation}%
This expression can be solved for $\lambda $ in terms of the the Lorentz
scalars $u^{2},\sigma ^{2},u\sigma \sigma u$ and $\det \sigma .$ But before
solving for $\lambda ,$ let us show that $\lambda $ and the lagrangian $L$
are related by the expression.

\begin{equation}
L=-m_{0}^{2}\lambda .  \tag{17}
\end{equation}

First using the constraints (10) and (11) the lagrangian (6) becomes

\[
L=u^{\mu }P_{\mu }+\frac{1}{2}\sigma ^{\mu \nu }\Sigma _{\mu \nu }. 
\]
From (7) and (9) we find the results

\begin{equation}
u^{\mu }P_{\mu }=\lambda P^{2}  \tag{18}
\end{equation}%
and

\begin{equation}
\frac{1}{2}\sigma ^{\mu \nu }\Sigma _{\mu \nu }=\lambda \frac{\Sigma ^{2}}{%
2r^{2}}.  \tag{19}
\end{equation}%
So, we get

\begin{equation}
L=u^{\mu }P_{\mu }+\frac{1}{2}\sigma ^{\mu \nu }\Sigma _{\mu \nu }=\lambda
\left( P^{2}+\frac{1}{2r^{2}}\Sigma ^{2}\right) .  \tag{20}
\end{equation}%
By using (11) once again, we see that (20) leads to (17).

Thus, (16) and (17) leads to the lagrangian

\begin{equation}
L=-m_{0}\left[ \frac{-\left( u^{2}+\frac{1}{2}r^{2}\sigma ^{2}\right) \pm 
\sqrt{\left( u^{2}+\frac{1}{2}r^{2}\sigma ^{2}\right) ^{2}+r^{2}\left(
u\sigma \sigma u-r^{2}\det \sigma \right) }}{2}\right] ^{1/2}.  \tag{21}
\end{equation}%
Observe that if $\sigma $ vanishes only the minus sign in the symbol $\pm 
\sqrt{}$ makes sense. In this case (21) is reduced to

\begin{equation}
L=-m_{0}\left[ -u^{2}\right] ^{1/2}  \tag{22}
\end{equation}%
which is the well known lagrangian for a relativistic point particle.

Since the lagrangian (21) is a function of the all Lorentz scalars that can
be formed from the velocities $u$ and $\sigma $, namely $u^{2},\sigma
^{2},u\sigma \sigma u$ and $\det \sigma $ we observe that such a lagrangian
has manifest Lorentz invariance. In fact, the lagrangian (21) has a Poincar%
\'{e} invariance under the infinitesimal transformations $\delta x^{\mu
}=a^{\mu }+\omega _{\nu }^{\mu }x^{\nu }$and $\delta e_{(\alpha )}^{\mu
}=\omega _{\nu }^{\mu }e_{(\alpha )}^{\nu },$ for arbitrary $\omega _{\mu
\nu }=-\omega _{\nu \mu }$. By explicit computation one can show applying
Noether's procedure to these transformations that $P^{\mu }$ and $M^{\mu \nu
}=x^{\mu }P^{\nu }-x^{\nu }P^{\mu }+\Sigma ^{\mu \nu }$ are conserved
generators obeying the Poincar\'{e} group algebra (see Refs. 3 and 4 for
details).

\bigskip\ 

\noindent \textbf{3.- THE CONSTRAINTS FROM THE SECOND ORDER LAGRANGIAN}

\smallskip\ 

The central idea in this section is to derive the set of constraints
associated with (21) which, of course, should correspond to (10) and (11).
Instead of starting with the lagrangian given in (21) we shall take
advantage of the formula (16) and (17). From this perspective one may assume
that $\lambda =\lambda \left( u,\sigma \right) $ and define a "linear
momentum"\ $p_{\mu }$ and an "internal angular momentum"\ $l_{\mu \nu }$ as 
\begin{equation}
p_{\mu }=\frac{\partial \lambda }{\partial u^{\mu }}  \tag{23}
\end{equation}%
and

\begin{equation}
l_{\mu \nu }=\frac{\partial \lambda }{\partial \sigma ^{\mu \nu }},  \tag{24}
\end{equation}%
respectively. According to (17) we have the relations: $P_{\mu
}=-m_{0}^{2}p_{\mu }$ and $\Sigma _{\mu \nu }=-m_{0}^{2}l_{\mu \nu }.$ One
of the reasons to define $p_{\mu }$ and $l_{\mu \nu }$ is to avoid carrying
all the time the factor $m_{0}^{2}$.

Taking the derivative of (16) with respect to $u^{\mu }$ leads to

\begin{equation}
\left[ \lambda ^{3}+\frac{\lambda }{2m_{0}^{2}}\left( u^{2}+\frac{1}{2}%
r^{2}\sigma ^{2}\right) \right] p_{\mu }=-\frac{\lambda ^{2}}{2m_{0}^{2}}%
u_{\mu }+\frac{r^{2}}{2m_{0}^{4}}u^{\alpha }\sigma _{\alpha \beta }\sigma
^{\beta }{}_{\mu }.  \tag{25}
\end{equation}

Let us define

\begin{equation}
A=\left[ \lambda ^{3}+\frac{\lambda }{2m_{0}^{2}}\left( u^{2}+\frac{1}{2}%
r^{2}\sigma ^{2}\right) \right] .  \tag{26}
\end{equation}%
We find

\begin{equation}
A^{2}p^{2}=\frac{\lambda ^{4}u^{2}}{4m_{0}^{4}}-\frac{\lambda ^{2}r^{2}}{%
2m_{0}^{6}}u\sigma \sigma u+\frac{r^{4}}{4m_{0}^{8}}u^{\alpha }\sigma
_{\alpha \beta }\sigma ^{\beta }{}_{\mu }\sigma ^{\mu }{}_{\tau }\sigma
^{\tau }{}_{\lambda }u^{\lambda }.  \tag{27}
\end{equation}%
But the identity (A19) leads to

\begin{equation}
u^{\alpha }\sigma _{\alpha \beta }\sigma ^{\beta }{}_{\mu }\sigma ^{\mu
}{}_{\tau }\sigma ^{\tau }{}_{\lambda }u^{\lambda }=u^{\alpha }\sigma
_{\alpha \beta }\left( -\frac{1}{2}\sigma ^{\beta }{}_{\lambda }\sigma ^{2}-%
\frac{1}{4}\sigma ^{\ast \beta }{}_{\lambda }\left( \sigma ^{\ast }\sigma
\right) \right) u^{\lambda }  \tag{28}
\end{equation}%
and therefore we have

\begin{equation}
A^{2}p^{2}=\frac{\lambda ^{4}}{4m_{0}^{4}}u^{2}-\frac{\lambda ^{2}r^{2}}{%
2m_{0}^{6}}u\sigma \sigma u-\frac{r^{4}}{8m_{0}^{8}}u\sigma \sigma u\sigma
^{2}+\frac{1}{64}\frac{r^{4}}{m_{0}^{8}}u^{2}\left( \sigma \cdot \sigma
^{\ast }\right) ^{2},  \tag{29}
\end{equation}%
where we used the fact that

\begin{equation}
\sigma _{\alpha \beta }\sigma ^{\ast \beta }{}_{\lambda }=-\frac{1}{4}\eta
_{\alpha \lambda }\left( \sigma \cdot \sigma ^{\ast }\right) .  \tag{30}
\end{equation}

Similarly, applying to (16) the derivative with respect to $\sigma ^{\mu \nu
}$ we obtain

\begin{equation}
\begin{array}{c}
\left[ \lambda ^{3}+\frac{\lambda }{2m_{0}^{2}}\left( u^{2}+\frac{1}{2}%
r^{2}\sigma ^{2}\right) \right] l_{\mu \nu }=-\frac{\lambda ^{2}r^{2}}{%
2m_{0}^{2}}\sigma _{\mu \nu } \\ 
\\ 
+\frac{r^{2}}{2m_{0}^{4}}\left( u^{\alpha }\sigma _{\alpha \mu }u_{\nu
}-u^{\alpha }\sigma _{\alpha \nu }u_{\mu }\right) +\frac{r^{4}}{8m_{0}^{4}}%
\sigma _{\mu \nu }^{\ast }\left( \sigma \cdot \sigma ^{\ast }\right) ,%
\end{array}
\tag{31}
\end{equation}%
where we used (A18). This expression yields

\begin{equation}
\begin{array}{c}
A^{2}l^{2}=\frac{\lambda ^{4}r^{4}}{4m_{0}^{4}}\sigma ^{2}-\frac{\lambda
^{2}r^{4}}{m_{0}^{6}}u\sigma \sigma u-\frac{\lambda ^{2}r^{6}}{8m_{0}^{6}}%
\left( \sigma \cdot \sigma ^{\ast }\right) ^{2} \\ 
\\ 
-\frac{r^{4}}{2m_{0}^{8}}u\sigma \sigma uu^{2}-\frac{r^{6}}{16m_{0}^{8}}%
u^{2}\left( \sigma \cdot \sigma ^{\ast }\right) ^{2}-\frac{r^{8}\sigma ^{2}}{%
64m_{0}^{8}}\left( \sigma \cdot \sigma ^{\ast }\right) ^{2},%
\end{array}
\tag{32}
\end{equation}%
where we used the identities $\sigma ^{\ast }\sigma ^{\ast }=-\sigma \sigma $
and (30).

Adding (29) and (32) leads to

\begin{equation}
\begin{array}{c}
A^{2}\left( p^{2}+\frac{1}{2r^{2}}l^{2}\right) =\frac{\lambda ^{4}}{%
4m_{0}^{4}}\left( u^{2}+\frac{1}{2}r^{2}\sigma ^{2}\right) -\frac{\lambda
^{2}r^{2}}{m_{0}^{6}}\left( u\sigma \sigma u-r^{2}\det \sigma \right) \\ 
\\ 
-\frac{r^{4}}{4m_{0}^{8}}\left( u^{2}+\frac{1}{2}r^{2}\sigma ^{2}\right)
\left( u\sigma \sigma u-r^{2}\det \sigma \right) ,%
\end{array}
\tag{33}
\end{equation}%
where we used (A18).

Using (16) and (26) we discover that

\begin{equation}
p^{2}+\frac{1}{2r^{2}}l^{2}=-\frac{1}{m_{0}^{2}}.  \tag{34}
\end{equation}%
Since $P_{\mu }=-m_{0}^{2}p_{\mu }$ and $\Sigma _{\mu \nu }=-m_{0}^{2}l_{\mu
\nu }$ we finally get from (34) the constraint

\begin{equation}
P^{2}+\frac{1}{2r^{2}}\Sigma ^{2}+m_{0}^{2}=0.  \tag{35}
\end{equation}

Let us now derive from (25) and (31) the constraint $\Sigma ^{\mu \nu
}P_{\nu }=0.$ We have

\begin{equation}
\begin{array}{c}
A^{2}l^{\mu \nu }p_{\nu }=\frac{\lambda ^{4}r^{2}}{2m_{0}^{2}}\sigma ^{\mu
\nu }u_{\nu }+\frac{\lambda ^{2}r^{2}u^{2}}{2m_{0}^{6}}\sigma ^{\mu \nu
}u_{\nu }-\frac{\lambda ^{2}r^{4}}{8m_{0}^{6}}(\sigma \cdot \sigma ^{\ast
})\sigma ^{\ast \mu \nu }u_{\nu } \\ 
\\ 
-\frac{\lambda ^{2}r^{4}}{2m_{0}^{6}}\sigma ^{\mu \nu }\sigma _{\nu \alpha
}\sigma ^{\alpha \beta }u_{\beta }-\frac{r^{4}(u\sigma \sigma u)}{2m_{0}^{8}}%
\sigma ^{\mu \nu }u_{\nu }+\frac{r^{4}u^{\mu }}{2m_{0}^{8}}u_{\tau }\sigma
^{\tau \nu }\sigma _{\nu \alpha }\sigma ^{\alpha \beta }u_{\beta } \\ 
\\ 
+\frac{r^{6}}{2m_{0}^{8}}(\sigma \cdot \sigma ^{\ast })\sigma ^{\ast \mu \nu
\tau \nu }\sigma _{\nu \alpha }\sigma ^{\alpha \beta }u_{\beta }.%
\end{array}
\tag{36}
\end{equation}%
The sixth term on the right hand side of (36) vanishes due to the identity $%
u_{\tau }\sigma ^{\tau \nu }\sigma _{\nu \alpha }\sigma ^{\alpha \beta
}u_{\beta }\equiv 0.$ Using (A19) and (30) it is not difficult to show that
(36) is reduced to

\begin{equation}
\begin{array}{c}
A^{2}l^{\mu \nu }p_{\nu }=\frac{\lambda ^{4}r^{2}}{2m_{0}^{2}}\sigma ^{\mu
\nu }u_{\nu }+\frac{\lambda ^{2}r^{2}}{2m_{0}^{6}}(u^{2}+\frac{r^{2}}{2}%
\sigma ^{2})\sigma ^{\mu \nu }u_{\nu } \\ 
\\ 
-\frac{r^{4}}{2m_{0}^{8}}(u\sigma \sigma u-r^{2}\det \sigma )\sigma ^{\mu
\nu }u_{\nu }.%
\end{array}
\tag{37}
\end{equation}%
Solving the factor $(u\sigma \sigma u-r^{2}\det \sigma )$ in (16) and
substituting the result in (37) we finally discover that $l^{\mu \nu }p_{\nu
}=0$. This leads to the constraint

\begin{equation}
\Sigma ^{\mu \nu }P_{\nu }=0,  \tag{38}
\end{equation}%
which is the Tulczyjew constraint.

Summarizing, we have shown that the lagrangian (21) leads to the constraints
(35) and (38) which were the starting point in the first order formalism of
section 2.

\bigskip\ 

\noindent \textbf{4.- FROM A POINT PARTICLE IN HIGHER DIMENSIONS TO THE DE
SITTER RELATIVISTIC TOP}

\smallskip\ 

Let us start writing the higher dimensional metric $\gamma _{MN}$ in terms
of the vielbien field $E_{M}^{A}$, 
\begin{equation}
\gamma _{MN}=E_{M}^{A}E_{N}^{B}\eta _{AB},  \tag{39}
\end{equation}%
where $\eta _{AB}$ is a \textit{flat metric} in $4+N$ dimensions. Here we
are considering the vielbien field $E_{M}^{A}$ as a the function of the
coordinates $x^{M}$.

The lagrangian of a point particle moving in a background determined by the
metric $\gamma _{MN}$ is

\begin{equation}
L=-M_{0}(-\gamma _{MN}\dot{x}^{M}\dot{x}^{N})^{1/2},  \tag{40}
\end{equation}%
where $M_{0}$ is the analogue of the mass of the object and $\dot{x}%
^{M}\equiv \frac{dx^{M}}{d\tau }$. From this lagrangian one gets the
Euler-Lagrange equations of motion

\begin{equation}
\frac{DP^{M}}{D\tau }\equiv \dot{P}^{M}+\Gamma _{RS}^{M}\dot{x}^{R}P^{S}=0, 
\tag{41}
\end{equation}%
where

\begin{equation}
P_{M}=\frac{\partial L}{\partial \dot{x}^{M}}  \tag{42}
\end{equation}%
and $\Gamma _{RS}^{M}$ are the Christoffel symbol associated with $\gamma
_{MN}$. We can write $E_{M}^{A}$ in the form

\begin{equation}
E_{M}^{A}=\left( 
\begin{array}{cc}
E_{\mu }^{a}\left( x,y\right) & E_{\mu }^{a^{\prime }}\left( x,y\right) \\ 
E_{i}^{a}\left( x,y\right) & E_{i}^{a^{\prime }}\left( x,y\right)%
\end{array}%
\right) ,  \tag{43}
\end{equation}%
where we used the notation $x^{M}=(x^{\mu },y^{i})=(x,y).$

Using the Kaluza-Klein mechanism it is well known that $E_{M}^{A}$ can be
written in the form

\begin{equation}
E_{M}^{A}=\left( 
\begin{array}{cc}
e_{\mu }^{a}\left( x\right) & \omega _{\mu }^{a^{\prime }}\left( x\right) \\ 
0 & e_{i}^{a^{\prime }}\left( y\right)%
\end{array}%
\right) .  \tag{44}
\end{equation}%
Here, for later convenience, we used the notation $\omega _{\mu }^{a^{\prime
}}\equiv E_{\mu }^{a^{\prime }},$ $e_{\mu }^{a}\equiv E_{\mu }^{a}$ and $%
e_{i}^{a^{\prime }}\equiv E_{i}^{a^{\prime }}.$

Using (44) we can brake the metric (39) in the form

\begin{equation}
\begin{array}{c}
\gamma _{\mu \nu }=g_{\mu \nu }+\omega _{\mu }^{a^{\prime }}\omega _{\nu
}^{b^{\prime }}\eta _{a^{\prime }b^{\prime }}, \\ 
\\ 
\gamma _{\mu i}=\omega _{\mu }^{a^{\prime }}e_{i}^{b^{\prime }}\eta
_{a^{\prime }b^{\prime }}, \\ 
\\ 
\gamma _{ij}=e_{i}^{a^{\prime }}e_{j}^{b^{\prime }}\eta _{a^{\prime
}b^{\prime }}=g_{ij}(y),%
\end{array}
\tag{45}
\end{equation}
where $g_{\mu \nu }(x)=e_{\mu }^{a}\left( x\right) e_{\nu }^{b}\left(
x\right) \eta _{ab}$.

The line element $ds^{2}$ associated with $\gamma _{MN}$ is

\begin{equation}
ds^{2}=\gamma _{MN}dx^{M}dx^{N}=\gamma _{\mu \nu }dx^{\mu }dx^{\nu }+2\gamma
_{\mu i}dx^{\mu }dy^{i}+\gamma _{ij}dy^{i}dy^{j}.  \tag{46}
\end{equation}%
Substituting (45) within (46) one gets

\begin{equation}
ds^{2}=\left( g_{\mu \nu }+\omega _{\mu }^{a^{\prime }}\omega _{\nu
}^{b^{\prime }}\eta _{a^{\prime }b^{\prime }}\right) dx^{\mu }dx^{\nu
}+2\left( \omega _{\mu }^{a^{\prime }}e_{i}^{b^{\prime }}\eta _{a^{\prime
}b^{\prime }}\right) dx^{\mu }dy^{i}+\left( e_{i}^{a^{\prime
}}e_{j}^{b^{\prime }}\eta _{a^{\prime }b^{\prime }}\right) dy^{i}dy^{j}. 
\tag{47}
\end{equation}%
This expression can be rewritten in the following form

\begin{equation}
ds^{2}=g_{\mu \nu }dx^{\mu }dx^{\nu }+\left( \omega _{\mu }^{a^{\prime
}}dx^{\mu }+e_{i}^{a^{\prime }}dy^{i}\right) ^{2},  \tag{48}
\end{equation}%
which is the Fukuyama starting point in the study of spinning particles.$%
^{17}$ If we choose the base$^{18-19}$

\begin{equation}
\theta ^{\mu }=dx^{\mu },  \tag{49a}
\end{equation}%
and

\begin{equation}
\theta ^{a^{\prime }}=\omega _{\mu }^{a^{\prime }}dx^{\mu }+e_{i}^{a^{\prime
}}dy^{i},  \tag{49b}
\end{equation}%
then we have that the line element (48) becomes

\begin{equation}
ds^{2}=g_{\mu \nu }\theta ^{\mu }\theta ^{\nu }+\eta _{a^{\prime }b}\theta
^{a^{\prime }}\theta ^{b^{\prime }}.  \tag{50}
\end{equation}%
Therefore, in the base (49) the metric takes the form

\begin{equation}
\hat{\gamma}_{MM}=\left( 
\begin{array}{cc}
g_{\mu \nu } & 0 \\ 
0 & \eta _{a^{\prime }b^{\prime }}%
\end{array}%
\right) .  \tag{51}
\end{equation}

The dual base is given by

\begin{equation}
D_{\mu }=\partial _{\mu }-\omega _{\mu }^{a^{\prime }}e_{a^{\prime
}}^{i}\partial _{i},  \tag{52a}
\end{equation}%
and

\begin{equation}
D_{a^{\prime }}=e_{a^{\prime }}^{i}\partial _{i}.  \tag{52b}
\end{equation}%
In fact, one can verify that

\begin{equation}
\left\langle \theta ^{\mu },D_{\nu }\right\rangle =\left\langle dx^{\mu
},\partial _{\nu }-\omega _{\nu }^{a^{\prime }}e_{a^{\prime }}^{i}\partial
_{i}\right\rangle =\delta _{\nu }^{\mu }  \tag{53}
\end{equation}%
and

\begin{equation}
\left\langle \theta ^{a^{\prime }},D_{b^{\prime }}\right\rangle
=\left\langle \omega _{\mu }^{a^{\prime }}dx^{\mu }+e_{i}^{a^{\prime
}}dx^{i},u_{b^{\prime }}^{j}\partial _{j}\right\rangle =e_{i}^{a^{\prime
}}e_{b^{\prime }}^{i}=\delta _{b^{\prime }}^{a^{\prime }}.  \tag{54}
\end{equation}%
Similarly, one can check that

\begin{equation}
\begin{array}{lll}
\left\langle \theta ^{a^{\prime }},D_{\nu }\right\rangle & = & \left\langle
\omega _{\mu }^{a^{\prime }}dx^{\mu }+e_{i}^{a^{\prime }}dx^{i},\partial
_{\nu }-\omega _{\nu }^{b^{\prime }}e_{b^{\prime \prime }}^{j}\partial
_{j}\right\rangle \\ 
&  &  \\ 
& = & \omega _{\nu }^{a^{\prime }}-e_{j}^{a^{\prime }}e^{jb^{\prime }}\omega
_{b^{\prime }\nu }=\omega _{\nu }^{a^{\prime }}-\eta ^{a^{\prime }b^{\prime
}}\omega _{b^{\prime }\nu }=0.%
\end{array}
\tag{55}
\end{equation}

Let us compute the commutator $\left[ D_{\mu },D_{\nu }\right] .$ From (52a)
we have

\begin{equation}
\begin{array}{lll}
\left[ D_{\mu },D_{\nu }\right] & = & \left[ \partial _{\mu }-\omega _{\mu
}^{a^{\prime }}e_{a^{\prime }}^{i}\partial _{i},\partial _{\nu }-\omega
_{\nu }^{b^{\prime }}e_{b^{\prime \prime }}^{j}\partial _{j}\right] \\ 
&  &  \\ 
& = & \partial _{\mu }\left( \partial _{\nu }-\omega _{\nu }^{b^{\prime
}}e_{b^{\prime \prime }}^{j}\partial _{j}\right) -\omega _{\mu }^{a^{\prime
}}e_{a^{\prime }}^{i}\partial _{i}\left( \partial _{\nu }-\omega _{\nu
}^{b^{\prime }}e_{b^{\prime \prime }}^{j}\partial _{j}\right) \\ 
&  &  \\ 
& = & -\partial _{\nu }\left( \partial _{\mu }-\omega _{\mu }^{a^{\prime
}}e_{a^{\prime }}^{i}\partial _{i}\right) +\omega _{\nu }^{b^{\prime
}}e_{b^{\prime \prime }}^{j}\partial _{j}\left( \partial _{\mu }-\omega
_{\mu }^{a^{\prime }}e_{a^{\prime }}^{i}\partial _{i}\right) .%
\end{array}
\tag{56}
\end{equation}

Considering that $[\partial _{\mu },\partial _{\nu }]=0,$ $[\partial
_{i},\partial _{j}]=0$ and $[\partial _{\mu },\partial _{i}]=0$ we find that
the expression (56) reduces to

\begin{equation}
\begin{array}{c}
\left[ D_{\mu },D_{\nu }\right] =\left( -\partial _{\mu }\omega _{\nu
}^{b^{\prime }}+\partial _{\nu }\omega _{\mu }^{b^{\prime }}\right)
e_{b^{\prime }}^{j}\partial _{j} \\ 
\\ 
+\left( \omega _{\mu }^{a^{\prime }}e_{a^{\prime }}^{i}\omega _{\nu
}^{b^{\prime }}-\omega _{\nu }^{c^{\prime }}e_{c^{\prime }}^{i}\omega _{\mu
}^{b^{\prime }}\right) \left( \partial _{i}e_{b^{\prime }}^{j}\right)
\partial _{j}.%
\end{array}
\tag{57}
\end{equation}%
The second term in (57) can be rewritten as

\begin{equation}
\left( \omega _{\mu }^{a^{\prime }}e_{a^{\prime }}^{i}\omega _{\nu
}^{b^{\prime }}-\omega _{\nu }^{c^{\prime }}e_{c^{\prime }}^{i}\omega _{\mu
}^{b^{\prime }}\right) \left( \partial _{i}e_{b^{\prime }}^{j}\right)
\partial _{j}=\omega _{\mu }^{a^{\prime }}\omega _{\nu }^{b^{\prime }}\left(
e_{a^{\prime }}^{i}\partial _{i}e_{b^{\prime }}^{j}-e_{b^{^{\prime }\prime
}}^{l}\partial _{l}e_{a^{\prime }}^{j}\right) .  \tag{58}
\end{equation}%
Let us write $e_{a^{\prime }}\equiv e_{a^{\prime }}^{i}\partial _{i},$ thus
we have

\begin{equation}
\left( e_{a^{\prime }}^{i}\partial _{i}e_{b^{\prime }}^{j}-e_{b^{^{\prime
}\prime }}^{l}\partial _{l}e_{a^{\prime }}^{j}\right) =\left[ e_{a^{\prime
}},e_{b^{\prime }}\right] .  \tag{59}
\end{equation}%
We assume that

\begin{equation}
\left[ e_{a^{\prime }},e_{b^{\prime }}\right] =-C_{a^{\prime }b^{\prime
}}^{d^{\prime }}e_{d^{\prime }},  \tag{60}
\end{equation}%
where $C_{a^{\prime }b^{\prime }}^{d^{\prime }}$ are the structure constants
associated with some group $G$. Substituting (60) into (57) we find the
expression

\begin{equation}
\left[ D_{\mu },D_{\nu }\right] =\left( -\partial _{\mu }\omega _{\nu
}^{b^{\prime }}+\partial _{\nu }\omega _{\mu }^{b^{\prime }}\right)
e_{b^{\prime }}-\omega _{\mu }^{a^{\prime }}\omega _{\nu }^{b^{\prime
}}C_{a^{\prime }b^{\prime }}^{d^{\prime }}e_{d^{\prime }},  \tag{61}
\end{equation}%
which by means of the definition

\begin{equation}
R_{\mu \nu }^{b^{\prime }}=\partial _{\mu }\omega _{\nu }^{b^{\prime
}}-\partial _{\nu }\omega _{\mu }^{b^{\prime }}+C_{c^{\prime }d^{\prime
}}^{b^{\prime }}\omega _{\mu }^{c^{\prime }}\omega _{\nu }^{d^{\prime }} 
\tag{62}
\end{equation}%
becomes

\begin{equation}
\left[ D_{\mu },D_{\nu }\right] =-R_{\mu \nu }^{b^{\prime }}e_{b^{\prime }}.
\tag{63}
\end{equation}%
Following similar procedure we find that

\begin{equation}
\left[ D_{\mu },D_{a^{\prime }}\right] =0  \tag{64}
\end{equation}%
and

\begin{equation}
\left[ D_{a^{\prime }},D_{b^{\prime }}\right] =-C_{a^{\prime }b^{\prime
}}^{d^{\prime }}e_{d^{\prime }}.  \tag{65}
\end{equation}%
Thus, from (63)-(65) we see that the only nonvanishing structure constants
related to the commutator $\left[ D_{M},D_{N}\right] $ are $C_{\mu \nu
}^{b^{\prime }}=-R_{\mu \nu }^{b^{\prime }}$ and $C_{a^{\prime }b^{\prime
}}^{d^{\prime }}$.

In general, in a non-base frame, the connection $\Gamma _{MNP}$ is given by 
\begin{equation}
\Gamma _{MNP}=\frac{1}{2}\left( D_{P}\hat{\gamma}_{MN}+D_{N}\hat{\gamma}%
_{MP}-D_{M}\hat{\gamma}_{NP}\right) +\frac{1}{2}\left(
C_{MNP}+C_{MPN}-C_{NPM}\right) .  \tag{66}
\end{equation}%
Since $C_{\mu \nu \alpha }=0$ and $\hat{\gamma}_{\mu \nu }=g_{\mu \nu }(x)$
we get

\begin{equation}
\Gamma _{\mu \nu \alpha }=\frac{1}{2}\left( D_{\alpha }\hat{\gamma}_{\mu \nu
}+D_{\nu }\hat{\gamma}_{\nu \alpha }-D_{\mu }\hat{\gamma}_{\nu \alpha
}\right) =\frac{1}{2}\left( \partial _{\alpha }g_{\mu \nu }+\partial _{\nu
}g_{\nu \alpha }-\partial _{\mu }g_{\nu \alpha }\right) \equiv \left\{ \mu
\nu \alpha \right\} .  \tag{67}
\end{equation}%
In the same way, since $C_{\mu \nu a^{\prime }}=-R_{\mu \nu a^{\prime }}$ we
obtain

\begin{equation}
\Gamma _{\mu \nu a^{\prime }}=\frac{1}{2}\left( C_{\mu \nu a^{\prime
}}+C_{\mu a^{\prime }\nu }-C_{\nu a^{\prime }\mu }\right) =-\frac{1}{2}%
R_{\mu \nu a^{\prime }}.  \tag{68}
\end{equation}%
We also get

\begin{equation}
\Gamma _{\mu a^{\prime }b^{\prime }}=0  \tag{69}
\end{equation}%
and

\begin{equation}
\Gamma _{a^{\prime }b^{\prime }c^{\prime }}=-\frac{1}{2}C_{a^{\prime
}b^{\prime }c^{\prime }}.  \tag{70}
\end{equation}

With these results in hand for $\Gamma _{MNP}$ we shall proceed to see their
consequences in the equations of motion (41). Let us start splitting (41) in
the following form

\begin{equation}
\dot{P}^{\mu }+\Gamma _{\nu \alpha }^{\mu }\dot{x}^{\nu }P^{\alpha }+\Gamma
_{\nu a^{\prime }}^{\mu }\dot{x}^{\nu }P^{a^{\prime }}+\Gamma _{a^{\prime
}\nu }^{\mu }\dot{x}^{a^{\prime }}P^{\nu }=0  \tag{71}
\end{equation}%
and

\begin{equation}
\dot{P}^{a^{\prime }}+\Gamma _{b^{\prime }c^{\prime }}^{a^{\prime }}\dot{x}%
^{b^{\prime }}P^{c^{\prime }}=0,  \tag{72}
\end{equation}%
where we used the fact that the only nonvanishing components of $\Gamma
_{MNP}$ are $\Gamma _{\mu \nu \alpha }$, $\Gamma _{\mu \nu a^{\prime }}$ and 
$\Gamma _{a^{\prime }b^{\prime }c^{\prime }}.$

Using (67) and (68) we discover that (71) and (72) yields

\begin{equation}
\frac{\bar{D}P^{\mu }}{\bar{D}\tau }=R_{\nu a^{\prime }}^{\mu }\dot{x}^{\nu
}P^{a^{\prime }}  \tag{73}
\end{equation}%
and

\begin{equation}
\dot{P}^{a^{\prime }}=0  \tag{74}
\end{equation}%
respectively, where $\frac{\bar{D}}{\bar{D}\tau }$ means covariant
derivative in terms of the Christoffel symbols $\left\{ \mu \nu \alpha
\right\} .$ Here we used the fact that

\begin{equation}
P_{M}=\frac{M_{0}\gamma _{MN}\dot{x}^{N}}{\left( -\gamma _{PQ}\dot{x}^{P}%
\dot{x}^{Q}\right) ^{\frac{1}{2}}},  \tag{75}
\end{equation}%
which means that $P^{M}=\lambda \dot{x}^{M}$, with $\lambda =M_{0}\left(
-\gamma _{PQ}\dot{x}^{P}\dot{x}^{Q}\right) ^{-\frac{1}{2}}.$

We shall show now that (73) and (74) are equivalent to the relativistic top
equations of motion (RTEM) in a gravitational field.$^{20-25}$ For that
purpose we shall make the indices identification $a^{\prime }\rightarrow
\left( a,b\right) $ where the pair $\left( a,b\right) $ is antisymmetric.
Thus, the equations of motion (73) and (74) become 
\begin{equation}
\frac{\bar{D}P^{\mu }}{\bar{D}\tau }=\frac{1}{2}R_{\nu ab}^{\mu }\dot{x}%
^{\nu }S^{ab}  \tag{76}
\end{equation}%
and

\begin{equation}
\dot{S}^{ab}=0,  \tag{77}
\end{equation}%
where we used the notation $S^{ab}\equiv P^{ab}$ and introduced in (76) the
quantity $\frac{1}{2}$ in order to avoid counting twice.

The last step is to write $S^{ab}=e_{\mu }^{a}e_{\nu }^{b}S^{\mu \nu }$ in
order to write (76) and (77) in the form

\begin{equation}
\frac{\bar{D}P^{\mu }}{\bar{D}\tau }=\frac{1}{2}R_{\nu \alpha \beta }^{\mu }%
\dot{x}^{\nu }S^{\alpha \beta }  \tag{78}
\end{equation}%
and

\begin{equation}
\frac{\bar{D}S^{\mu \nu }}{\bar{D}\tau }=0,  \tag{79}
\end{equation}%
which are the traditional forms given to the RTEM in a gravitational field.

It is interesting to clarify the meaning of the constant $M_{0}.$ From (75)
it follows the constraint

\begin{equation}
P^{M}P^{N}\hat{\gamma}_{MM}=-M_{0}^{2}.  \tag{80}
\end{equation}%
which in virtue of the form of the metric $\hat{\gamma}_{MM},$ given in
(51), we see that (80) can be written as

\begin{equation}
g_{\mu \nu }P^{\mu }P^{\nu }+\eta _{a^{\prime }b^{\prime }}P^{a^{\prime
}}P^{b^{\prime }}=-M_{0}^{2}  \tag{81}
\end{equation}%
or

\begin{equation}
g_{\mu \nu }P^{\mu }P^{\nu }+\frac{1}{2r^{2}}S^{\mu \nu }S_{\mu \nu
}=-m_{0}^{2},  \tag{82}
\end{equation}%
where we used the relation $S^{ab}=e_{\mu }^{a}e_{\nu }^{b}S^{\mu \nu }$ and
redefined $P^{\mu }$ as $rP^{\mu }$ and $M_{0}$ as $rm_{0}$ with $r$ a
constant of the motion. If we compare the expression (82) with (1) we
observe their great similarity. However, they are not exactly the same
because the constraint $\Sigma ^{\mu \nu }P_{\nu }\thickapprox 0$ given in
(4) it is not satisfied by $P^{\mu }$ and $S^{\mu \nu }.$ Instead of the
constraint (4) we can define the vector

\begin{equation}
S^{\mu }=\frac{1}{2}\epsilon ^{\mu \nu \alpha \beta }P_{\nu }S_{\alpha \beta
}  \tag{83}
\end{equation}%
and, as a consequence of this formula, we have

\begin{equation}
S^{\mu }P_{\mu }=0.  \tag{84}
\end{equation}%
Nevertheless, the relation between $\Sigma ^{\mu \nu }$ and $S^{\mu \nu }$
is subtle and requires a careful analysis. First of all, let us write the
first order lagrangian

\begin{equation}
L=\dot{x}^{M}P_{M}-\frac{\lambda ^{\prime }}{2}(P^{M}P_{M}+M_{0}^{2}), 
\tag{85}
\end{equation}%
corresponding to (40). Using (82) we see that (40) can be written as

\begin{equation}
L=\dot{x}^{\mu }P_{\mu }+\frac{1}{2}\dot{x}^{\mu \nu }S_{\mu \nu }-\frac{%
\lambda }{2}(P^{\mu }P_{\mu }+\frac{1}{2r^{2}}S^{\mu \nu }S_{\mu \nu
}+m_{0}^{2}),  \tag{86}
\end{equation}%
where $\dot{x}^{\mu \nu }=e_{a}^{\mu }e_{b}^{\nu }\dot{x}^{ab}$ and $\lambda
\equiv r^{2}\lambda ^{\prime }.$ Comparing (86) with (6) one observes the
close similarity between both lagrangians. We can even try to go from (86)
to (6) by using the transformation

\begin{equation}
S^{\mu \nu }=\Sigma ^{\mu \nu }+\xi ^{\mu }P^{\nu }-\xi ^{\nu }P^{\mu }. 
\tag{87}
\end{equation}%
But this implies to redefine the velocities $u^{\mu }$ and $\sigma ^{\mu \nu
}$ in terms of the velocities $\dot{x}^{\mu }$ and $\dot{x}^{\mu \nu }.$
This is related to the fact that the motion of a relativistic top can be
described, in an equivalent way, by two position vectors, namely, the center
of mass and the center of charge. (An extensive discussion about the meaning
of these two position vectors of a relativistic top can be found in
reference 26.) The center of mass can be associated with $\Sigma ^{\mu \nu }$
via the constraint $\Sigma ^{\mu \nu }P_{\nu }\thickapprox 0$, while the
center of charge with $S^{\mu \nu }.$ Thus, the transformation (87) suggests
to identify the variable $\xi ^{\mu }$ with the difference between the
center of mass and the center of charge. In fact, multiplying (87) by $%
P_{\mu }$ one gets

\begin{equation}
S^{\mu \nu }P_{\nu }=\xi ^{\mu }P^{\nu }P_{\nu },  \tag{88}
\end{equation}%
where we assumed $\xi ^{\mu }P_{\mu }=0.$ Substituting this expression into
(87) one discovers that $\Sigma ^{\mu \nu }$ can be obtained from $S^{\mu
\nu }$ using the projector $\eta ^{\mu \alpha }-\frac{1}{P^{2}}P^{\mu
}P^{\alpha }.$

Summarizing, we have shown that using Kaluza-Klein theory it is possible to
obtain the relativistic top theory from a point particle in higher
dimensions. This is in fact a very interesting result because it means that
although the top does not follow geodesics in four dimensions, it does in
higher dimensions.

\bigskip\ 

\noindent \textbf{5.- FINAL COMMENTS}

\smallskip\ 

In this article we have shown different aspects of a particular relativistic
top, namely the top satisfying (1) and (4). First, we showed the equivalence
between the first and second order lagrangians (6) and (21). Then, the form
of the lagrangian (6) motivated us to look for a higher dimensional
description of the top, and as a matter of fact we discovered that it is
possible to obtain the relativistic top equations of motion starting from a
geodesic equation of motion of a point particle in higher dimensions. This
is an interesting result that deserves to be analyzed in terms of a fiber
bundle scenario.

First we notice that such a result is similar to the case of the Lorentz
force associated with a charged particle which can be obtained from a
geodesic in five dimensions. More generally, our result is similar to the
generalized Lorentz force associated with a Yang-Mills gauge field which can
be obtained by a geodesic in $4+D$-dimensions. In this case, the traditional
method is to consider a $4+D$-dimensional principle fiber bundle $P,$ which
locally looks like $M^{4}\times B$, where $M^{4}$ is a four-dimensional base
space and $B$ is a group manifold whose dimension is $D$. The key object to
connect the geodesic in $4+D$-dimensions with the generalized Lorentz force
in four dimensions is the one-form in the cotangent space $T^{\ast }(P),$

\begin{equation}
\omega =g^{-1}dg+g^{-1}Ag,  \tag{89}
\end{equation}%
where $A=A_{\mu }^{a}T_{a}dx^{\mu }$ can eventually be identified with the
Yang-Mills gauge field. Here, $T_{a}$ are the generators of some group $G$
acting transitively on $B$ and having the properties

\begin{equation}
\lbrack T_{a},T_{b}]=C_{ab}^{c}T_{c}.  \tag{90}
\end{equation}

In principle, if we consider the fiber space $M^{4}\times Q,$ where $Q$
corresponds also to a 4-dimensional manifold, one may apply similar
description to the case of a spinning object (see Ref. 17). In this case the
connection one-form reads as

\begin{equation}
\omega =g^{-1}dg+g^{-1}\Omega g,  \tag{91}
\end{equation}%
where $\Omega $ is given by

\begin{equation}
\Omega =\frac{1}{2}\omega _{\mu }^{AB}S_{AB}dx^{\mu }.  \tag{92}
\end{equation}%
Here, we shall assume that $S_{AB}$ are the generators of the de Sitter
group $SO(1,4)$ (or anti de Sitter group $SO(2,3)$). If we compare (49b)
with (91) we observe that both expressions are very similar. In fact, since $%
g$ is an element of $SO(1,4)$ we can write $g$ as a matrix in the form $%
\Lambda _{B}^{A}$ and therefore the one-form (91) yields

\begin{equation}
\omega ^{AB}=\omega _{i}^{AB}dy^{i}+\omega _{\mu }^{AB}dx^{\mu },  \tag{93}
\end{equation}%
where $\omega _{i}^{AB}=\Lambda ^{CA}\partial _{i}\Lambda _{C}^{B}$. This
expression can be written as

\begin{equation}
\omega ^{5a}=\omega _{i}^{5a}dy^{i}+\omega _{\mu }^{5a}dx^{\mu }  \tag{94}
\end{equation}%
and

\begin{equation}
\omega ^{ab}=\omega _{i}^{ab}dy^{i}+\omega _{\mu }^{ab}dx^{\mu }.  \tag{95}
\end{equation}%
The base (49a) and (49b) arises from (94) and (95) by defining $%
e_{i}^{a^{\prime }}\equiv \omega _{i}^{ab}$, $e_{\mu }^{a}\equiv \omega
_{\mu }^{5a}$ and setting $\omega _{i}^{5a}$ equal to zero. This means that
we can write (44) in the following form

\begin{equation}
\omega _{M}^{AB}=\left( 
\begin{array}{cc}
\omega _{\mu }^{5a}\left( x\right) & \omega _{\mu }^{ab}\left( x\right) \\ 
0 & \omega _{i}^{ab}\left( y\right)%
\end{array}%
\right) .  \tag{96}
\end{equation}%
Therefore, the metric $\gamma _{MN}$ in (39) becomes

\begin{equation}
\gamma _{MN}=\frac{1}{2}\omega _{M}^{AB}\omega _{NAB}  \tag{97}
\end{equation}%
and consequently the lagrangian (40) can be written as

\begin{equation}
L=-M_{0}(-\frac{1}{2}\omega _{M}^{AB}\omega _{NAB}\dot{x}^{M}\dot{x}%
^{N})^{1/2},  \tag{98}
\end{equation}%
Thus, the corresponding line element is

\begin{equation}
ds^{2}=\frac{1}{2}\omega _{M}^{AB}\omega _{NAB}dx^{M}dx^{N},  \tag{99}
\end{equation}%
which is in agreement with the Fukuyama's suggestion.$^{17}$ We shall call
the system described by the lagrangian (98) with $\omega _{M}^{AB}$ given by
(96) the \textit{de Sitter top.}

There are several observations that one can make from our analysis but
perhaps one of the most important is the fact that the top theory leads
naturally to consider the de Sitter group $SO(1,4)$ (or anti de Sitter group 
$SO(2,3)$) via the connection $\omega _{M}^{AB}$. As it is well known, this
group structure appears in several fronts of current interest, including
Maldacena's conjecture in string theory, accelerated universe, and the
gravitational gauge approach of MacDowell-Mansouri. For this reason it turns
out to be of particular interest for further research to study the possible
connection between the de Sitter top and these lines of research.

On the other hand, since in the Kaluza-Klein $4+N+D$-dimensional space the
metric gives gravity, Yang-Mills, and scalar fields, a top moving in this
space will be acted on by these fields. It may be interesting to see how the
Yang-Mills field, which can be understood as the field which generates
rotations in the isospace, affects the motion of the top. In the Ph. D.
thesis in reference 27 some computations in this direction were reported. We
consider convenient to attach a brief review of such computations as an
appendix (see appendix B). Part of our task, for further research, is to
analyze these computations from the point of view of the present work.

Finally, the Regge trajectory (1) which lead us to the de Sitter
relativistic top concept is a very particular relation between the mass and
the spin of a particle. In general, a singular lagrangian can lead to more
general Regge trajectories of the form (B5). In particular, by using group
theoretical methods Atre and Mukunda$^{7}$ have develop a systematic
procedure for other possible Regge trajectories. In those cases one should
expect generalizations or alternatives of the de Sitter relativistic top
which is based in the de Sitter group. From the group theoretical point of
view, however, such a generalizations or alternatives appears as an
intriguing possibilities motivating further research on the subject.

\bigskip\ 

\noindent \textbf{APPENDIX A}

\smallskip\ 

In this appendix we present the proof of formula (15). From (7) and (10) we
get

\begin{equation}
u^{2}=\lambda ^{2}P^{2}+V^{2}.  \tag{A1}
\end{equation}%
Similarly, from (9), (10) and (12) we find

\begin{equation}
\sigma ^{2}=\frac{\lambda ^{2}}{r^{4}}\Sigma ^{2}+2\xi ^{2}P^{2}.  \tag{A2}
\end{equation}%
Here we also used the notation $b^{2}=b^{\mu \nu }b_{\mu \nu },$ for any
tensor $b^{\mu \nu }$.

We shall compute

\begin{equation}
\sigma ^{4}\equiv \sigma _{\mu \nu }\sigma ^{\nu }{}_{\alpha }\sigma
^{\alpha }{}_{\beta }\sigma ^{\beta \mu }.  \tag{A3}
\end{equation}%
Observe first that

\begin{equation}
\sigma ^{\mu \alpha }\sigma _{\alpha }{}^{\nu }=\frac{\lambda ^{2}}{r^{4}}%
\Sigma ^{\mu \alpha }\Sigma _{\alpha }{}^{\nu }-\frac{\lambda }{r^{2}}\left(
V^{\mu }P^{\nu }+V^{\nu }P^{\mu }\right) -\xi ^{2}P^{\mu }P^{\nu }-P^{2}\xi
^{\mu }\xi ^{\nu },  \tag{A4}
\end{equation}%
where we used (10) and (12). The expression (A4) leads to 
\begin{equation}
\sigma ^{4}=\frac{\lambda ^{4}}{r^{8}}\Sigma ^{4}+4\frac{\lambda ^{2}}{r^{4}}%
P^{2}V^{2}+2\xi ^{2}\xi ^{2}P^{2}P^{2},  \tag{A5}
\end{equation}%
where once again we used (10) and (12) and antisymmetric properties, such as 
$\Sigma ^{\mu \nu }\xi _{\mu }\xi _{\nu }\equiv 0.$

We also need to compute

\begin{equation}
u\sigma \sigma u\equiv u_{\mu }\sigma ^{\mu \nu }\sigma _{\nu }{}^{\alpha
}u_{\alpha }.  \tag{A6}
\end{equation}%
From (7), (9), (10) and (12) we find

\begin{equation}
\sigma ^{\mu \nu }u_{\nu }=\frac{\lambda }{r^{2}}\Sigma ^{\mu \nu }V_{\nu
}-\lambda P^{2}\xi ^{\mu }.  \tag{A7}
\end{equation}%
Therefore, we have 
\begin{equation}
u\sigma \sigma u=-\lambda ^{2}\left( -\frac{1}{r^{4}}V_{\mu }\Sigma ^{\mu
\nu }\Sigma _{\nu }{}^{\alpha }V_{\alpha }+\frac{2}{r^{2}}%
P^{2}V^{2}+P^{2}P^{2}\xi ^{2}\right)  \tag{A8}
\end{equation}%
or

\begin{equation}
u\sigma \sigma u=-\lambda ^{2}\left( \frac{1}{r^{4}}\xi _{\alpha }\Sigma
^{\alpha \nu }\Sigma _{\nu \tau }\Sigma ^{\tau \lambda }\Sigma _{\lambda
}{}^{\beta }\xi _{\beta }+\frac{2}{r^{2}}P^{2}V^{2}+P^{2}P^{2}\xi
^{2}\right) .  \tag{A9}
\end{equation}

Now, let us define the dual of any antisymmetric tensor $A_{\alpha \beta }$
as 
\begin{equation}
A^{\ast \mu \nu }=\frac{1}{2}\epsilon ^{\mu \nu \alpha \beta }A_{\alpha
\beta },  \tag{A10}
\end{equation}%
where $\epsilon ^{\mu \nu \alpha \beta }$ is the completely antisymmetric
Levi-Cevita tensor.

It turns out that from the constraint (10) it follows

\begin{equation}
\Sigma ^{\ast \mu \nu }\Sigma _{\mu \nu }=0.  \tag{A11}
\end{equation}%
Using (A11) it is not difficult to show that

\begin{equation}
\Sigma ^{\mu \nu }\Sigma _{\nu \alpha }\Sigma ^{\alpha \tau }=-\frac{1}{2}%
\Sigma ^{\mu \tau }\left( \Sigma ^{2}\right) .  \tag{A12}
\end{equation}%
Thus, using (A12) one finds that (A9) becomes

\begin{equation}
u\sigma \sigma u=-\lambda ^{2}\left( \frac{1}{2r^{4}}V^{2}\Sigma ^{2}+\frac{2%
}{r^{2}}P^{2}V^{2}+P^{2}P^{2}\xi ^{2}\right) .  \tag{A13}
\end{equation}

On the other hand from (A2) we obtain

\begin{equation}
\sigma ^{2}\sigma ^{2}=\frac{\lambda ^{4}}{r^{8}}\Sigma ^{2}\Sigma ^{2}+%
\frac{4\lambda ^{2}}{r^{4}}\xi ^{2}P^{2}\Sigma ^{2}+4\xi ^{2}\xi
^{2}P^{2}P^{2}.  \tag{A14}
\end{equation}%
Therefore, (A5) and (A14) imply%
\begin{equation}
\sigma ^{2}\sigma ^{2}-2\sigma ^{4}=\frac{4\lambda ^{2}}{r^{4}}\xi
^{2}P^{2}\Sigma ^{2}-\frac{8\lambda ^{2}}{r^{4}}P^{2}V^{2},  \tag{A15}
\end{equation}%
where we used the fact that from (A12) it follows that

\begin{equation}
2\Sigma ^{4}-\Sigma ^{2}\Sigma ^{2}=0.  \tag{A16}
\end{equation}

Now, define

\begin{equation}
\det \sigma =\frac{1}{4!}\varepsilon ^{\mu \nu \alpha \beta }\varepsilon
^{\tau \lambda \sigma \rho }\sigma _{\mu \lambda }\sigma _{\nu \lambda
}\sigma _{\alpha \sigma }\sigma _{\beta \rho }.  \tag{A17}
\end{equation}%
It is not difficult to show that

\begin{equation}
\det \sigma =-\frac{1}{16}\left( \sigma ^{\ast }\sigma \right) ^{2}. 
\tag{A18}
\end{equation}

From the identity

\begin{equation}
\sigma ^{\mu \alpha }\sigma _{\alpha \beta }\sigma ^{\beta \nu }=-\frac{1}{2}%
\sigma ^{\mu \nu }\left( \sigma ^{2}\right) -\frac{1}{4}\sigma ^{\ast \mu
\nu }\left( \sigma ^{\ast }\sigma \right) ,  \tag{A19}
\end{equation}%
we find

\begin{equation}
\sigma ^{4}=\frac{1}{2}\sigma ^{2}\sigma ^{2}+\frac{1}{4}\left( \sigma
^{\ast }\sigma \right) \left( \sigma ^{\ast }\sigma \right) .  \tag{A20}
\end{equation}%
Therefore, by combining (A18) and (A20) we obtain

\begin{equation}
\det \sigma =\frac{1}{8}\left( \sigma ^{2}\sigma ^{2}-2\sigma ^{4}\right) . 
\tag{A21}
\end{equation}

From (A15) we see that (A21) implies

\begin{equation}
\det \sigma =\frac{\lambda ^{2}}{2r^{4}}\xi ^{2}P^{2}\Sigma ^{2}-\frac{%
\lambda ^{2}}{r^{4}}P^{2}V^{2}.  \tag{A22}
\end{equation}%
Now, from (A13) and (A22) we see that 
\begin{equation}
\begin{array}{c}
u\sigma \sigma u-r^{2}\det \sigma =\lambda ^{2}\left[ \frac{-1}{2r^{4}}%
V^{2}\Sigma ^{2}-\frac{2}{r^{2}}P^{2}V^{2}-\xi ^{2}r^{2}P^{2}-\frac{1}{2r^{2}%
}\xi ^{2}P^{2}\Sigma ^{2}+\frac{1}{r^{2}}P^{2}V^{2}\right] \\ 
\\ 
=\lambda ^{2}\left[ \frac{m_{0}^{2}}{r^{2}}V^{2}+m_{0}^{2}\xi ^{2}P^{2}%
\right] .%
\end{array}
\tag{A23}
\end{equation}%
Using (A1) and (A2) one finally sees that (A23) leads to (15).

\bigskip\ 

\noindent \textbf{APPENDIX B}

\smallskip\ 

In $4+N^{\prime }$dimensions, with $N^{\prime }=N+D,$ a top can be described
by the variables $x^{M}(\tau )$ and $E_{A}^{M}(\tau )$, where $x^{M}$ are $%
4+N^{\prime }$ coordinates, $E_{A}^{M}(\tau )$ are $4+N^{\prime }$
orthonormal vectors, $\tau $ is an arbitrary parameter and the index $A$ in
parenthesis labels the name of the vector . The vectors $E_{A}^{M}$ satisfy
the condition%
\begin{equation}
\gamma _{MN}=E_{M}^{A}E_{N}^{B}\eta _{AB},  \tag{B1}
\end{equation}%
where $\eta _{AB}=diag\left( -1,1,....,1\right) $ is an scalar matrix and $%
\gamma _{MN}$ is the curved metric generalized to $4+N^{\prime }$ dimensions.

Define the linear velocity $u^{M}$ and the angular velocity $\sigma ^{MN}$
as follows:

\begin{equation}
u^{M}\equiv \frac{dx^{M}}{d\tau }=\dot{x}^{M},  \tag{B2}
\end{equation}

\begin{equation}
\sigma ^{MN}=\eta ^{AB}E_{A}^{M}\frac{\Delta }{d\tau }E_{B}^{N}=-\sigma
^{NM}.  \tag{B3}
\end{equation}%
Here, the symbol $\frac{\Delta }{d\tau }$ means covariant derivative with
respect to $\tau $, having the Christoffel symbols $\Gamma _{NP}^{M}$ as the
connection. One sees that $\sigma ^{MN}$ is again antisymmetric by virtue of
the condition $\left( B1\right) .$

Consider a top with linear velocity $u^{M}$, angular velocity $\sigma ^{MN},$
linear momentum $P^{M}$, and internal angular momentum $S^{MN}$. We will
assume that the dynamics of the system is generated by the lagrangian

\begin{equation}
L=-u^{M}P_{M}-\frac{1}{2}\sigma ^{MN}S_{MN}+\lambda H+\lambda _{M}H^{M}, 
\tag{B4}
\end{equation}%
where

\begin{equation}
H\equiv P^{M}P_{M}-f\left( \frac{1}{2}S^{MN}S_{MN\;}\right)  \tag{B5}
\end{equation}%
and

\begin{equation}
H^{M}\equiv S^{MN}P_{N}  \tag{B6}
\end{equation}%
corresponds to the Regge and Tulcyzjew constraints respectively, generalized
to $4+N^{\prime }$ dimensions. Here, $\lambda $ and $\lambda _{M}$ are
lagrange multipliers.

Using the lagrangian (B4) and assuming the Equivalence Principle in $%
4+N^{\prime }$ dimensions leads to the RTEM equations of motion generalized
to higher dimensions,

\begin{equation}
\frac{\Delta P^{M}}{d\tau }=-\frac{1}{2}R_{NPQ}^{M}u^{N}S^{PQ}  \tag{B7}
\end{equation}%
and

\begin{equation}
\frac{\Delta S^{MN}}{d\tau }=P^{M}u^{N}-P^{N}u^{M}.  \tag{B8}
\end{equation}%
Here, $\frac{\Delta A^{M}}{d\tau }=\frac{dA^{M}}{d\tau }+\Gamma
_{NP}^{M}A^{N}u^{P},$where $A^{M}$ is any vector and $R_{NPQ}^{M}$ is the
Riemann tensor.

Using the constraint $H^{M}=0$ one sees from (B8) that $J^{2}=\frac{1}{2}%
S^{MN}S_{MN}$ is a constant of the motion. Using this fact and the
constraint $H=0$ one can show that $P^{M}P_{M}=-M^{2}$ is also a constant of
the motion.

We will follow the strategy of doing the computations in the \textit{%
horizontal lift base }defined by the commutators;$^{18-19}$

\[
\left[ D_{\mu },D_{\nu }\right] =-F_{\mu \nu }\text{ }^{a}D_{a}, 
\]

\begin{equation}
\left[ D_{\mu },D_{a}\right] =0,  \tag{B9}
\end{equation}

\[
\left[ D_{a},D_{b}\right] =f_{ab}\text{ }^{c}D_{c}. 
\]%
Therefore the only nonvanishing commutation coefficients are

\begin{equation}
C_{\mu \nu }\text{ }^{c}=-F_{\mu \nu }\text{ }^{c}  \tag{B10}
\end{equation}%
and

\begin{equation}
C_{ab}\text{ }^{c}=f_{ab}\text{ }^{c}.  \tag{B11}
\end{equation}%
In the base (B9) the metric associated with $M^{4}\times N^{\prime }$ is

\begin{equation}
\hat{\gamma}_{MN}=\left( 
\begin{array}{cc}
g_{\mu \nu } & 0 \\ 
0 & g_{ab}%
\end{array}%
\right) ,  \tag{B12}
\end{equation}%
with

\[
D_{a}g_{\mu \nu }=0, 
\]

\[
D_{c}g_{ab}=f_{cab}+f_{cba}, 
\]

\begin{equation}
D_{\mu }g_{ab}=\partial _{\mu }g_{ab}-A_{\mu }^{c}D_{c}g_{ab}=g_{ab\mid \mu
},  \tag{B13}
\end{equation}

\[
D_{\alpha }g_{\mu \nu }=g_{\mu \nu },_{\alpha }. 
\]

The Christoffel symbols are given in a non-coordinate base by

\[
\hat{\Gamma}_{MNP}=\frac{1}{2}\left( D_{P}\hat{\gamma}_{MN}+D_{N}\hat{\gamma}%
_{MP}-D_{N}\hat{\gamma}_{NP}\right) 
\]

\begin{equation}
+\frac{1}{2}\left( C_{MNP}+C_{MPN}-C_{NPM}\right) ,  \tag{B14}
\end{equation}%
and the Riemann tensor is given by

\begin{equation}
\hat{R}_{NPQ}^{M}=D_{P}\hat{\Gamma}_{NQ}^{M}-D_{Q}\hat{\Gamma}_{NQ}^{M}+\hat{%
\Gamma}_{RP}^{M}\hat{\Gamma}_{NQ}^{R}-\hat{\Gamma}_{RQ}^{M}\Gamma _{NP}^{R}-%
\hat{\Gamma}_{NR}^{M}C_{PQ}^{R}.  \tag{B15}
\end{equation}

The linear velocity $u^{M}$, angular velocity $\sigma ^{MN},$ linear
momentum $P^{M}$,and internal angular momentum $S^{MN}$ will be referred
below with respect to the "horizontal lift base"\ defined in (B9). By using
(B12)-(B15) one may reduce the equations of motion (B7) to four dimensions;

\begin{equation}
\frac{D\Pi ^{\mu }}{d\tau }=-\frac{1}{2}^{4}R_{\nu \alpha \beta }^{\mu
}u^{\nu }S^{\alpha \beta }+Q_{a}F_{\nu }^{\mu }u^{\nu }+\frac{1}{4}F_{\alpha
\beta }\text{ }^{a;\mu }M^{\alpha \beta }\text{ }_{a}+Z^{\mu }.  \tag{B16}
\end{equation}%
Here the following definitions were used:

\begin{equation}
\Pi ^{\mu }\equiv P^{\mu }-\frac{1}{2}g_{ab}F_{\nu }^{\mu }\text{ }%
^{a}S^{\nu b},  \tag{B17}
\end{equation}

\begin{equation}
Q_{a}\equiv g_{ab}P^{b}+\frac{1}{4}g_{ab}f_{ce}\text{ }^{b}S^{ce}+\frac{1}{4}%
g_{ab}F_{\alpha \beta }\text{ }^{b}S^{\alpha \beta }+\frac{1}{2}g_{ab\mid
\alpha }S^{\alpha b},  \tag{B18}
\end{equation}

\begin{equation}
M^{\alpha \beta }\text{ }_{a}\equiv g_{ab}u^{a}S^{\alpha \beta
}+g_{ab}u^{\beta }S^{\alpha b}-g_{ab}u^{\alpha }S^{\beta b},  \tag{B19}
\end{equation}%
and

\[
Z^{\mu }=[\left( \frac{1}{2}u^{a}P^{b}-\frac{1}{4}F_{\alpha \beta }\text{ }%
^{a}u^{\alpha }S^{\beta b}+\frac{1}{4}F_{\alpha \beta }\text{ }^{a}S^{\alpha
\beta }u^{b}-\frac{1}{4}g^{af}g_{fe\mid \alpha }u^{e}S^{\alpha b}\right) 
\]

\[
-\frac{1}{4}g^{af}g_{fe\mid \alpha }u^{\alpha }S^{eb}+\left( f_{ec}\text{ }%
^{a}+f_{c}^{a}\text{ }_{e}+f_{e}^{a}\text{ }_{c}\right) u^{e}S^{cb} 
\]

\begin{equation}
-f_{ec}\text{ }^{b}u^{a}S^{ec})]g_{ab}\text{ }^{\mid \mu }+\frac{1}{2}g_{ab}%
\text{ }^{\mid \mu }\text{ };_{\alpha }S^{\alpha b}u^{a}.  \tag{B20}
\end{equation}%
Here, the symbol $g_{ab}$ $^{\mid \mu };_{\alpha }$ means

\[
g_{ab}\text{ }^{\mid \mu };_{\alpha }=g_{ab}\text{ }^{\mid \mu }\text{ }%
_{\mid \alpha }+\left\{ _{\beta \alpha }^{\mu }\right\} g_{ab}\text{ }^{\mid
\beta }, 
\]%
where $g_{ab}$ $^{\mid \mu }$ is defined in (B13)$.$ While the symbol $%
F_{\alpha }^{\mu }$ $^{a};_{\beta }$ means

\[
F_{\alpha }^{\mu a};_{\beta }=F_{\alpha }^{\mu a},_{\beta }+\left\{ _{\sigma
\beta }^{\mu }\right\} F_{\alpha }^{\sigma }\text{ }^{a}-\left\{ _{\beta
\alpha }^{\sigma }\right\} F_{\sigma }^{\mu }\text{ }^{a}. 
\]

In terms of the definitions (B17) and (B19) the equation of motion (B9)
leads to

\begin{equation}
\frac{DS^{\mu \nu }}{d\tau }=\Pi ^{\mu }u^{\nu }-\Pi ^{\nu }u^{\mu }+\frac{1%
}{2}\left( F^{\mu \alpha a}M_{\alpha }^{\nu }\text{ }_{a}-M^{\mu \alpha }%
\text{ }_{a}F_{\alpha }^{\nu }\text{ }^{a}\right) +H^{\mu \nu },  \tag{B21}
\end{equation}%
where

\[
H^{\mu \nu }=S^{\mu a}u^{b}g_{ab}\text{ }^{\mid \nu }-S^{\nu a}u^{b}g_{ab}%
\text{ }^{\mid \mu }. 
\]%
Clearly, equations (B16) and (B21) are generalizations of the usual
4-dimensional case. One can show that the quantity $Q_{a}$ given in (B18) is
a constant of the motion. It turns out that $Q_{a}$\ can be interpreted as
charges of the system. For details the reader is referred to Ref. 27.

\bigskip\ 

\noindent \textbf{Acknowledgments: }We would like to thank M. C. Mar\'{\i}n
and for her helpful comments.

\end{document}